\documentclass[a4paper,fleqn]{cas-dc}

\usepackage[numbers]{natbib}
\usepackage{amsmath,amssymb,amsfonts}
\usepackage{algorithmic}
\usepackage{graphicx}
\usepackage{textcomp}
\usepackage{xcolor}
\usepackage{comment}
\usepackage{array}
\usepackage{multirow}
\usepackage{makecell}
\usepackage{url}
\usepackage{breakurl}
\usepackage{pifont}
\usepackage{balance}
\usepackage{hyperref}
\usepackage{adjustbox}
\usepackage{caption}
\usepackage{soul}
\newcommand{\colorTextMod}{\textcolor{Black}}
\newcommand{\cmark}{\ding{51}}%
\newcommand{\xmark}{\ding{55}}%

\def\tsc#1{\csdef{#1}{\textsc{\lowercase{#1}}\xspace}}
\tsc{WGM}
\tsc{QE}
\tsc{EP}
\tsc{PMS}
\tsc{BEC}
\tsc{DE}

\begin{document} \sloppy
\let\WriteBookmarks\relax
\def\floatpagepagefraction{1}
\def\textpagefraction{.001}
\shorttitle{Cryptomining Makes Noise}
\shortauthors{Caprolu et~al.}

\title [mode = title]{Cryptomining Makes Noise: a Machine Learning Approach for Cryptojacking Detection}                      



\author[1]{Maurantonio Caprolu}[type=editor,
                        auid=000,bioid=1,
                        orcid=0000-0001-7511-2910]
\cormark[1]
\ead{mcaprolu@mail.hbku.edu.qa}


\address[1]{Hamad Bin Khalifa University (HBKU)  \\College of Science and Engineering (CSE), \\Division of Information and Computing Technology (ICT).  \\Doha, Qatar}

\author[1]{Simone Raponi}[type=editor,
                        auid=000,bioid=1,
                        orcid=0000-0001-7511-2910]
\ead{sraponi@mail.hbku.edu.qa}

\author[1]{Gabriele Oligeri}[type=editor,
                        auid=000,bioid=1,
                        orcid=0000-0001-7511-2910]
\ead{goligeri@mail.hbku.edu.qa}

\author[1]{Roberto {Di Pietro}}[type=editor,
                        auid=000,bioid=1,
                        orcid=0000-0001-7511-2910]
\ead{rdipietro@hbku.edu.qa}

\cortext[cor1]{Corresponding author}


\begin{abstract}
\colorTextMod{A new cybersecurity attack,where an adversary illicitly runs crypto-mining software over the devices of unaware users, is emerging in both the literature and in the wild . This attack,  known as cryptojacking, has proved to be very effective given the simplicity of running a crypto-client into a target device. Several countermeasures have recently been proposed, with different features and performance, but all characterized by a host-based architecture. 
This kind of solutions, designed to protect the individual user, are not suitable for efficiently protecting a corporate network, especially against insiders.\\*
In this paper, we propose a network-based approach to detect and identify crypto-clients activities by solely relying on the network traffic, even when encrypted. First, we provide a detailed analysis of the real network traces generated by three major cryptocurrencies, Bitcoin, Monero, and Bytecoin, considering both the normal traffic and the one shaped by a VPN. Then, we propose Crypto-Aegis, a Machine Learning (ML) based framework built over the results of our investigation, aimed at detecting cryptocurrencies related activities, e.g., pool mining, solo mining, and active full nodes. Our solution 
achieves a striking 0.96 of F1-score and 0.99 of AUC for the ROC, while enjoying a few other properties, such as device and infrastructure independence. 
 Given the extent and novelty of the addressed threat we believe that our approach, supported by its excellent results, 
 pave the way for further research in this area.}
\end{abstract}




\maketitle

\section{Introduction}

Blockchain actually tries to solve the old problem of \emph{distributed consensus} by exploiting solutions matured from decades of research~\cite{Haber}. The solution coming from Bitcoin's blockchain is particularly interesting: entities participating to the ``voting" process should prove to have solved a moderately hard puzzle, the so called Proof-of-Work (PoW)~\cite{Tschorsch}. Indeed, for the vast majority of cryptocurrencies, in order to verify a transaction and to
have it added to the distributed ledger, 
participants are requested to compute a PoW.
Computationally solving PoW is referred as \emph{mining}. Over time, the complexity of puzzle solving (typically based on hashing as per Bitcoin and several others, such as Altcoin) has increased, 
leading to a rush for deploying more and more powerful systems that nowadays are able to compute more than $40 \cdot 10^{18}$ hashes per second (worldwide hash rate for Bitcoin at the time of writing this paper\footnote{https://www.blockchain.com/en/charts/hash-rate}). 
ASIC architectures are today guaranteeing the best trade-off between power consumption, terrific hash rate, size, cost, and life-time. The recent adoption of ASIC architectures brings in again the major issue of centralization~\cite{monero_pow}. Indeed, the huge gap between CPU/GPU and ASIC mining makes the latter the only viable way to participate to the network as a miner. Eventually, this causes centralization since only ASIC-based crypto-miners can participate to the consensus process. In order to mitigate the above trend, other digital currencies have been created that are actually exploiting different PoW strategies, being therefore ASIC-resistant. For instance, 
\emph{Monero} is an example of cryptocurrency specifically designed to be mined also with CPU-based architectures. Indeed, \emph{Monero} adopts the \emph{CryptoNight} PoW algorithm where 
the marginal benefit derived from specialized architectures such as GPU, FPGA, or ASIC does not introduce any significant gain for justifying the adoption of such a hardware. Therefore, mining could also be profitable when performed via CPU-based architectures. The \emph{CryptoNight} algorithm works by filling a segment of cache with random data corresponding to memory addresses, then subsequently hashing the resulting block after reading and writing to those addresses~\cite{monero_hash}.

PoW is becoming a significant source of revenues for the entities participating to the consensus process. This phenomena will grow even more with the increasing number of users joining the digital currency markets. However, 
while PoW computational requirements are  fueling methodologies and techniques to achieve more and more computational power with less energy consumption, new malicious practices involves PoW-offloading to unaware users. 
Indeed, a very recent cybersecurity attack involves the illicit use of resources from an unaware users to carry out PoW, i.e., \emph{cryptojacking}~\cite{8406561}. This attack mainly consists on the unauthorized mining of cryptocurrencies allowing malicious parties to steal resources in terms of CPU, GPU, and memory from a target machine with the aim of effortlessly collecting crypto-wealth. This behaviour is gaining momentum for two main reasons: the ease of deployment of crypto-clients; and, the difficulty to detect those crypto-clients.  
While being a general threat, cryptojacking is becoming particularly critical in Corporate ICT where the vast majority of laptops, desktops, and smartphones are distributed among the employees under a limited (if any) supervision. Indeed,
several unauthorized mining activities have already been discovered. Russian nuclear scientists have been arrested for ``Bitcoin mining plot''~\cite{russia_nuclear_mining}, the US government banned a Professor for secretly mining with National Science Foundation supercomputers~\cite{National_Science_Foundation}, a former Federal Reserve employee was sentenced to 12 months probation and a \$5,000 fine after pleading guilty to installing unauthorized software that connected to an online Bitcoin network in order to earn units of the digital currency~\cite{Federal_Reserve}, a Harvard student used 14,000-Core supercomputer to mine Dogecoin~\cite{Harvard_Student}, the factory lines of Hoya, the leading manufacturer of optical products in Japan, were shut down for three days as hackers tried to establish an unauthorized cryptocurrency mining~\cite{hoya}, just to cite a few.\\

{\bf Contribution.} The major contributions of this paper are listed below:
\begin{itemize}
    \item{\em A new type of attack}: we define a novel type of attack that subsumes the cryptojacking attack, i.e., the \emph{sponge-attack}, where  an adversary (either internal or external) secures a personal profit illicitly exploiting third party computing resources.
    \colorTextMod{\item{\em Network traffic analysis}: we provide a detailed analysis of real network traffic generated by 3 major cryptocurrencies: Bitcoin, Monero, and Bytecoin.
    \item{\em Encrypted traffics}: we investigate how VPN tunneling shapes the network traffic generated by Crypto-clients by considering two major VPN brands: NordVPN and ExpressVPN;}
    \item{\em The Crypto-Aegis Framework}: We propose Crypto-Aegis, a Machine Learning (ML) based framework built over the previous steps to detect crypto-mining activities; 
    \colorTextMod{\item{\em Comparison}: We compare our results against competing solutions in the literature.}
\end{itemize}

Crypto-Aegis enjoys the following features: (i) {\em Infrastructure independence}. The analysis is performed at the exit points (edge) of the Corporate network, independently of network size, network layout, and even when multiple layers of encryption are set in place by the attacker, e.g., a VPN is in use; (ii) {\em Device Independence}. We do not require any modification to the already existing devices adopted by the Corporate employees; (iii) {\em Multi-adversarial profiles support}. Our solution detects the presence of illicit behaviours via  network traffic analysis and independently of the adversarial profiles, i.e., be it an insider or an outsider;  (iv) {\em No clean state required}. Our solution detects the presence of a miner independently of the time the miner started its activities; and, (v)  {\em Effectiveness}. Our solution achieves an F1-score of 0.96 and the AUC of the ROC is greater than 0.99.\\

{\bf Roadmap.} The paper is organized as follows. Section~\ref{sec:related} resumes the most important contributions \colorTextMod{related to both cryptojacking analysis and detection. Section~\ref{sec:background} provides some background concepts related to the ML tools used in our solution, and presents the most important ML techniques for network traffic classification.} 
Section~\ref{sec:scenario} introduces the scenario and the adversary model. The details related to the measurement setup are depicted in Section~\ref{sec:measurement_setup}, while a throughout analysis of the collected network traffic traces is presented in Section~\ref{sec:network_traffic}. Section~\ref{sec:traffic_classification} depicts a baseline example analysis, i.e., Bitcoin vs standard software, introducing all the statistics that will be considered for the subsequent analysis. Sections~\ref{sec:detection_full_nodes} and~\ref{sec:detection_miners} introduce the metodologies used by our Crypto-Aegis framework, related to the detection and identification of full nodes and miners, respectively. Finally, Section~\ref{sec:sponge_attack_detection} tackles with the general problem of detecting a Crypto-node in a Corporate network, while a detailed discussion of our results and a comparison with other solutions from the literature is presented in Section~\ref{sec:discussion}. Section~\ref{sec:conclusion} draws some concluding remarks.

\section{Related Work}
\label{sec:related}

The computational power required to validate and to add blocks to the Bitcoin blockchain has greatly limited the odds that individuals without specialized hardware can provide any contribution to this process. Dedicating small devices (e.g., smartphones, laptops, desktop), or more powerful ones (e.g., workstations, servers), to the mining process would not be worth the cost of the electricity. This discourages users and leaves only a few in the world the opportunity to make contributions and earn the rewards arising. With the advent of other CPU-based cryptocurrencies this scenario has undergone many changes. History repeats itself again. In other ages, by seeing a mine populated by mechanical diggers, the gold digger with the only pick-axe on his shoulders would be forced to find new promising shores. This return to the ``gold rush'' led to the rediscovery of numerous attacks that had lost meaning with Bitcoin. This type of attacks are identified by the term \emph{cryptojacking}. Hackers, as well as dishonest employees who would like to round off their earnings, ``borrow'' resources belonging to others to run the mining process. Hand in hand with threats, some solutions have been already proposed, with the aim of implementing countermeasures to mitigate their effects.
\colorTextMod{\subsection{Cryptojacking Analysis}}
In~\cite{zimba2018crypto}, the state-of-the art of crypto-mining attacks have been investigated. By analyzing the malware code, as well as its behavior upon execution, authors examine two common attacks: web browser-based crypto-mining, and installable binary crypto-mining, respectively. Browser-based crypto-mining attacks exploit the JavaScript technology of web-pages, leveraging two web technology's advancements: \texttt{asm.js} and \texttt{WebAssembly}~\cite{konoth2018minesweeper}. Installable binary crypto-mining instead, is possible by using modified versions of the \emph{XMrig} software~\cite{xmrig}. The paper analyzes the techniques adopted by cybercriminals to establish a persistence mechanism and avoid detection, and it introduces both static and dynamic analysis, useful to uncover the techniques employed by the malware to exploit potential victims.
In~\cite{hong2018you}, the authors present an in-depth study over cryptojacking. The analysis of 853,936 popular web pages led to the identification of 2770 unique cryptojacking samples, of which 868 belonging to Alexa's top 100k ranking websites. A similar solution has been proposed by~\cite{musch2018web}. The authors propose an approach aiming to identify mining scripts, conducting a large-scale study on the prevalence of cryptojacking in the Alexa's 1 million websites. According to the analysis, on average 1 out of 500 websites hosts a mining script. Numerous works have followed the same direction. In~\cite{8406561}, authors conduct measurements to establish the cryptojacking relevance and profitability, wondering whether it should be classified as an attack or as a business opportunity. In~\cite{ruth2018digging}, 138 million domains have been explored, of which 137 million among com/net/org domains and 1 million coming from the Alexa's Top 1 million list. The analysis shows that the prevalence of browser mining is currently the 0.08\% of the analyzed set, a worrying number that should not be underestimated. The Alexa's Top 1 million websites have been taken into account even by~\cite{konoth2018minesweeper}, in which the authors studied the websites affected by drive-by mining to understand the techniques being used to evade the detection. As a result, 20 active crypto-mining campaigns have been identified. In \cite{musch2018web} \colorTextMod{the authors proposed a 3-phase analysis approach to investigate the cryptojacking phenomenon. They conducted a large-scale study on the Alexa first 1 million websites, finding that approximately 1 out of 500 sites hosts a mining script that immediately starts mining activities when visited.}

\colorTextMod{\subsection{Cryptojacking Detection}}
\colorTextMod{One of the first methodologies used to identify cryptojacking was the analysis of static signatures, as typically done for other types of malware~\cite{ALAEIYAN201976}. Several solutions, such as~\cite{drmine} and~\cite{MinerBlock}, implement static methods to detect mining activities and blacklist malicious web sites. This approach has been proved ineffective against cryptojacking, because of the usage of  obfuscation techniques to evade detection~\cite{CMBlock_Razali19}}.
A first step towards the application of Machine Learning techniques \colorTextMod{to cryptojacking detection} has been made by~\cite{carlin2018detecting}. The authors present an experimental study in which the dynamic opcode analysis successfully allows the browser-based crypto-mining detection. The proposed model can distinguish among crypto-mining sites, weaponized benign sites (e.g., benign sites to which the crypto-mining code has been injected), de-weaponized crypto-mining sites (e.g., crypto-mining sites to which the \texttt{start()} call has been removed), and real world benign sites. 
In~\cite{liu2018novel}, the authors presented a method to detect the browser's malicious mining behavior. Heap snapshot and stack features have been asynchronously extracted and automatically classified using Recurrent Neural Networks (RNNs). With 1159 malicious samples analyzed, the experimental results show that the proposed prototype recognizes the original mining samples with 98\% of accuracy if not encrypted, 93\% otherwise. 
In~\cite{hong2018you}, \colorTextMod{after identifying a set of inherent characteristics of cryptojacking scripts, such as the repeated hash-based computations and the regular call stack, a behavior-based detector called \emph{CMTracker} has been introduced. In} \cite{8737381} \colorTextMod{the authors proposed CapJack, a machine learning-based detection mechanism able to spot in-browser malicious cryptocurrency mining activities. This solution leverages CapsNet, a machine learning algorithm that mimic biological neural organization. CapJack makes use of system features such as CPU, Memory, disk and network utilization, implementing an host-based solution with a detection rate of 87\%. Another approach has been used in}~\cite{CMBlock_Razali19}\colorTextMod{, where the authors propose an application browser extension, named as CMBlock, able to detect and block mining scripts contained on web pages. The proposed solution combines two different methodologies: blacklisting and a mining behaviour detection technique.\\
All the solutions described in this section are host-based countermeasures designed to detect a mining script running on a single host. For this reason, these solutions are not able to effectively protect a corporate network from the cryptojacking threats, as described in Section~\ref{sec:scenario}. In fact, host-based solutions should be installed in every host belonging to the corporate network, with high installation and maintenance costs, not to mention privacy issues. Besides, these solutions use computational resources of the host that they must protect, subtracting them from the business tasks they should be  dedicated to.  
}

\colorTextMod{\section{Background}\label{sec:background}}
\colorTextMod{This section provides the reader with background knowledge on the most important techniques used in this paper. The first part contains a description of the Machine Learning techniques that have been adopted in the proposed solution. Then, in the second part we introduce related work that have laid the foundations for the classification of network traffic using Machine Learning algorithms.}

\colorTextMod{\subsection{Machine Learning Tools}
\textbf{Random Forest.} Random Forest}~\cite{breiman2001random} \colorTextMod{is an ensemble supervised Machine Learning technique, built as a combination of tree predictors. As an ensemble learning technique, the classification is the result of a decision taken collectively, from a large number of classifiers. The idea behind ensembles classification is based upon the premise that a set of classifiers can provide a more accurate and generalized (thus, less prone to overfitting) classification than a single classifier. With Random Forest, each classifier is a tree, and each tree depends on the values of a random vector independently sampled, but with the same distribution for all the trees in the forest. In detail, for the $i^{th}$ tree, a random vector
$\theta_i$ is generated, independent of the past random vectors $\theta_1$,...,$\theta_{i-1}$, but with the same distribution. Each tree \textit{i} grows using the training set and $\theta_i$, resulting in a classifier h(x, $\theta_i$), where $x$ in an input vector. After a sufficiently large number of trees is generated, each tree casts a unit vote for the most popular class at input $x$. }
\colorTextMod{To guarantee a degree of diversity among the base decision trees, a randomization approach is used, which works well both with bagging and random subspace methods}~\cite{kulkarni2013random}. \colorTextMod{To generate each single tree with the Random Forest algorithm, several steps are involved. Let $N$ be the number of records in the training set, and $M$ be the number of input variables. The training set (also known as bootstrap sample) is built by sampling $N$ records at random with replacement from the original data. At each node, $m$ variables (with $m << M$) are randomly selected out of $M$. The best split of these $m$ attributes is used to split the node. Once the forest is built, a new instance will run across all the trees in the forest. Each tree provides a classification for the new instance and issues a vote. The majority of the votes will allow to objectively declare the result of the new instance's classification.}\\
\colorTextMod{\textbf{k-Fold Cross-Validation.} k-Fold Cross-Validation}~\cite{kohavi1995study} \colorTextMod{is an accuracy estimation method that allows to evaluate how the results of a model will be generalized to an independent dataset. The main objective of cross-validation methods is to estimate the generalization of a model, that is, to understand its accuracy in the classification of data that it had never seen before (i.e., to avoid the overfitting problem). The method consists in partitioning the dataset into subsets, some of which (e.g., training set) will be used to perform the training of the model, while the remaining ones will be used for validation (e.g., validation set) or for testing (e.g., testing set) purposes. There are two types of cross-validation methods: exhaustive cross-validation methods and non-exhaustive cross-validation methods, respectively. The only difference is the number of subsets generated for the split that are performed. In fact, while the exhaustive cross-validation methods use all possible splitting combinations for training and testing, non-exhaustive methods use only a subset of them. We can therefore say that the non-exhaustive methods are an approximation of the exhaustive ones. K-fold cross-validation is an instance of non-exhaustive methods. In k-fold cross-validation the dataset $D$ is randomly split into $k$ mutually exclusive subsets: $D_1, D_2$, \dots, $D_k$ of approximately equal size. The model is trained and tested $k$ times, in particular each time $t \in \{1, 2, \dots, l\}$ the model is trained on $D - D_t$ and tested on $D_t$. The cross-validation estimate of accuracy is the number of correct classifications divided by the number of instances in the dataset}~\cite{kohavi1995study}.

\colorTextMod{\subsection{Machine Learning Techniques for Network Traffic Classification.} Network traffic classification has gained more and more attention in the very recent years, having the potential to solve several problems in network management~\cite{LU201660, SANVITO2019175} (e.g., building network profiles for proactive real time network traffic monitoring and management), as well as in network security~\cite{WANG201915} (e.g., Machine Learning application for intrusion detection systems or anomaly detection).}

\colorTextMod{The classic approach combines the analysis of the packets header with the payload inspection. Despite the high accuracy of this methodology, the high volume  of data to be  processed, together with the users privacy issues implied by this approach,  pushed the research community to explore different techniques. Moreover,  payload inspection is not possible in case of encrypted traffic, that nowadays is practically the norm. 
A promising research direction explores Machine Learning techniques for both real time IP traffic classification and static offline analysis of previously captured traffic.
One of the first work that allowed to understand the effectiveness of Machine Learning algorithms for the classification of network traffic is~}\cite{zander2005automated}. \colorTextMod{The authors make use of unsupervised Machine Learning techniques to automatically classify traffic flows based on statistical flow characteristics. After studying and evaluating the influence of each feature, including Forward-Pkt-Len-Var, Backward-Pkt-Len-Var, Backward-Bytes, Forward-Pkt-Len-Mean, Forward-Bytes, Backward-Pkt-Len-Mean, Duration, and Forward-IAT-Mean, several traffic traces collected at different locations of the Internet have been used to evaluate the efficiency of the adopted approach.}
In~\cite{nguyen2008survey}, \colorTextMod{the authors survey significant Machine Learning-based IP traffic classification solutions proposed in the literature. They highlight that the use of different Machine Learning algorithms for offline traffic analysis (e.g., AutoClass, Expectation Maximization, Decision Tree, Naive Bayes) provides high accuracy (up to 99\%) for different Internet applications traffic.} In~\cite{soysal2010machine}, \colorTextMod{authors evaluate different Machine Learning algorithms for flow-based network traffic classification, in terms of correctness and computational cost. In particular, they investigate the use of three supervised algorithms (i.e., Bayesian Networks, Decision Trees and Multilayer Perceptrons) considering six different classes: Peer-to-Peer (P2P), web (HTTP), content delivery (Akamai), bulk (FTP), service (DNS) and mail (SMTP). Their results show that Decision Trees have both a higher accuracy and a higher classification rate than Bayesian Networks. However, Decision Trees require a larger build time and are more susceptible in the case of incorrect or small amounts of training data. Moreover, they highlight that the amount of training data for a certain traffic class can affect the classification accuracy of both itself and other traffic classes. For this reason, they propose a systematic approach to construct specific training sets that feature the best accuracy results.
Regarding the analysis of encrypted network traffic, an early solution is from }~\cite{alshammari2009machine}. \colorTextMod{The authors use different Machine Learning algorithms (i.e., AdaBoost, Support Vector Machine, Naive Bayesian, RIPPER, and C4.5) to distinguish SSH traffic from non-SSH traffic in a given traffic trace. Their results show that the model generated by C4.5 algorithm outperforms the other ones using flow-based features only.}
\colorTextMod{Another contribution in this field is given by}~\cite{conti2016analyzing} and~\cite{conti2015can}. \colorTextMod{By using supervised Machine Learning techniques for Android encrypted network traffic analysis, the authors demonstrate that an external attacker can identify the specific actions that a user is performing on his mobile apps. Using a Random Forest classifier, they are able to infer not only the app used by the target user, but also the specific action he performed (e.g., sending an e-mail, posting a message, refreshing the home, and so on) for the most used Android applications, such as Gmail, Facebook, and Twitter, despite the use of SSL/TLS for traffic encryption.}

\section{Scenario, Assumptions, and Adversary Model}
\label{sec:scenario}
In the following, we describe our reference scenario, the assumptions we make as for the network infrastructure and network traffic classification and, finally, the adversary model.
\subsection{Scenario} 
\begin{figure}[]
\includegraphics[width=\columnwidth]{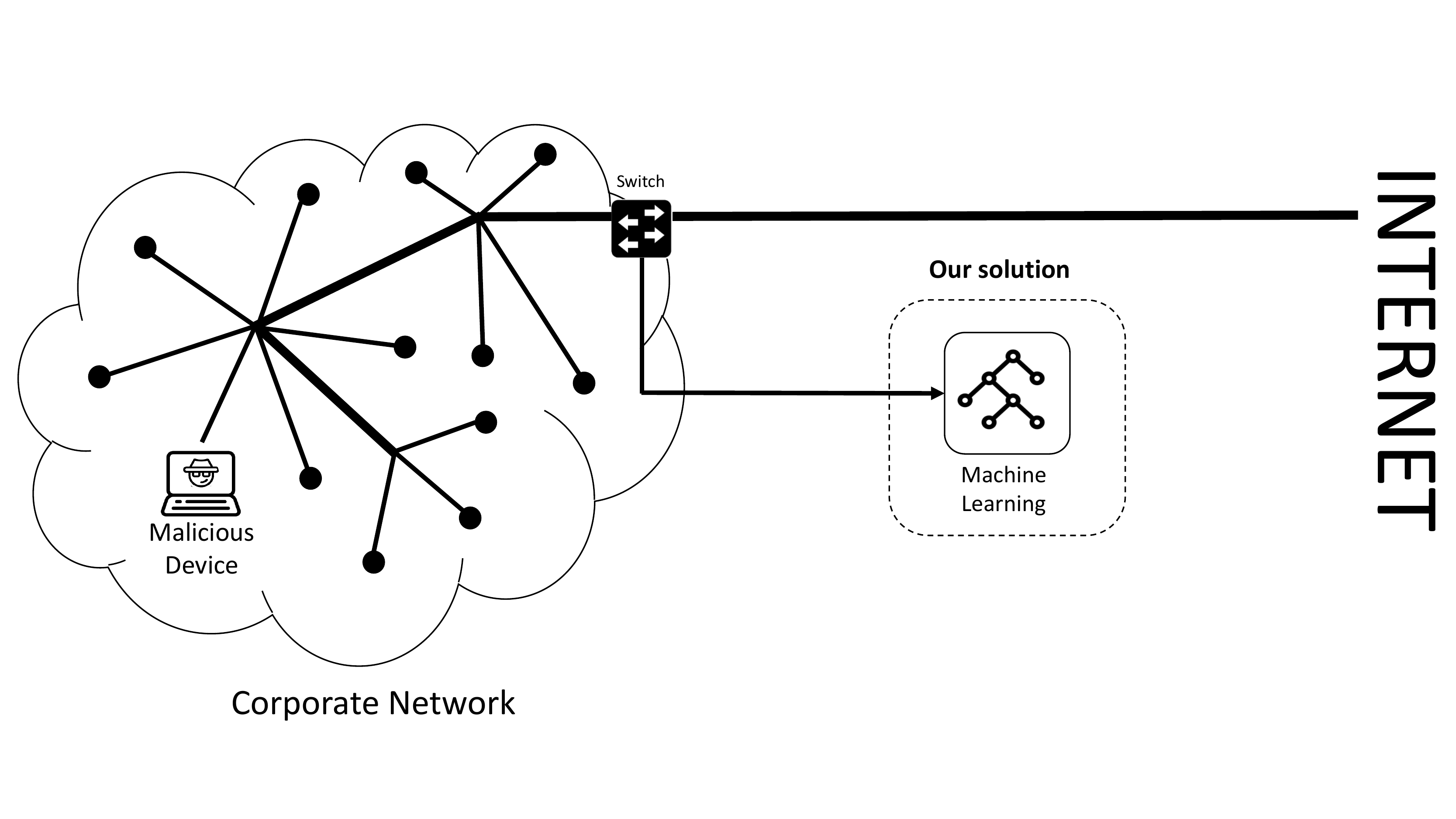}
\centering
\caption{Network scenario: A Corporate device mines cryptocurrencies controlled by a malicious entity. Crypto-Aegis is constituted by a Machine Learning algorithm classifying the traffic coming from an Ethernet switch at the edge of the Corporate Network.}
\label{fig:scenario}
\end{figure}

Figure~\ref{fig:scenario} shows the details of our reference scenario. 
We consider a corporate network constituted by several interconnected devices, including one that  is controlled by a malicious entity, willing to mine cryptocurrencies without being detected. Our solution should be deployed at the network edge and it involves only an Ethernet connection from the main Corporate Network switch to a server running our Machine Learning algorithm. We observe that our solution  requires interventions neither on the employee devices, nor on the already existing network infrastructure. Moreover, our solution can be easily deployed even when there are multiple exit connections between the Corporate Network and the Internet: this can be easily achieved by deploying multiple Ethernet links to collect the data from the Corporate Network exit points.
We observe that the above configuration is very conservative with respect to standard commercial solutions. Indeed, in the vast majority of cases, corporate solutions involve hardware for deep packet inspection deployed before the exit point, or even at multiple locations of the network. On the one hand  the association between traffic and device is much easier, while on the other hand  it requires a significant cost in terms of hardware equipment and deployment. In our solution, we consider the traffic already aggregated, i.e., affected by IP masquerading/NAT, or even tunneled and re-encrypted by a Virtual Private Network (VPN).

\subsection{Adversary model}
\label{sec:adversarial_model}

We consider two adversary models with respect to the corporate network: (i) \emph{insider}; and, (ii) {\em outsider}. We assume the insider has direct access to the hardware resources of the company, and therefore, has the opportunity to install new software into it. A typical example is the employee willing to accumulate crypto-wealth by exploiting corporate resources such as CPU, GPU, and network bandwidth. Moreover, we envisage an external adversary (outsider) being able to inject one or more corporate devices with a malicious software for performing unauthorized crypto-mining. Typical examples might be both the increasing number of malware delivering crypto-mining software to unaware users, and websites running crypto-mining Java scripts without the user's consent. Our adversary model (as depicted in Fig.~\ref{fig:scenario}) takes into account a corporate device illicitly running crypto-mining-related activities. As previously stated, we stress that our model takes into account a malicious device that might be controlled by either a dishonest employee or by a remote hacker who took over control of the device itself. 

There are several strategies to mine without the company consent, an activity that today is really difficult to detect and prevent. This malicious behaviour might be implemented by either a \emph{full node} or a \emph{miner}. 

\begin{itemize}
    \item{\em Full node}. It is a full-featured client of the cryptocurrency infrastructure. It locally stores the whole blockchain and participates to the consensus algorithm, being able to validate all the transactions. The mining activity performed by a full node is called \textit{solo mining} because the process is done independently from other nodes.
    \item{\em Miner}. It is a lightweight software that implements a simple worker that receives jobs (i.e., hash computations useful for the PoW) from a third party (i.e., a mining pool). When the mining pool successfully mines a block, both the reward and the fees will be divided among all the participants, proportionally to the computational power offered. This software does not participate in the cryptocurrency protocol and does not require to store a blockchain to work.
\end{itemize}

In our scenario, we assume that the client (being either a full node or a miner) is already provided with the ledger, if needed, and does not require any warm-up operations. Indeed, Crypto-Aegis does not resort to application-specific transients and it does not require to be deployed before the malicious device starts its illicit activity.
\\\\
{\bf \em Definition.} We define \emph{sponge-attack} as the malicious behavior of exploiting third-party hardware and software resources to obtain a personal profit without the authorization of the infrastructure's owner. The sponge-attack illicitly absorbs resources from the targeted infrastructure and makes a pay off out of them in favor of the attacker.\\\\
This definition is more general than the one of Cryptojacking, that only refers to unauthorized mining activities. The sponge-attack, instead, includes also any other activities performed with someone else's resources without authorization. An example of sponge-attack could be a malicious full node installed on a corporate server to perform a DDoS attack against a cryptocurrency network by using the company's network resources.

The sponge-attack can be implemented by deploying either a Full Node or a Miner. \\
{\bf Miner.} The use of a mining pool software allows to carry out mining activities without installing the heavier full node software. An adversary can use this software to perform a faster and stealthier attack since the targeted device does not need to store the ledger, usually very large. Furthermore, by joining a mining pool, profits are increased even if the available resources are limited. 
\\
{\bf Full node.} Deploying a full node into a network without the administrator consent has significant advantages for the adversary. Firstly, the full node gives to the adversary the capability to perform \textit{solo mining}, if the victim's resources are sufficiently powerful. Moreover, the full node could be used to attack the cryptocurrency's network, by performing double spending attacks, DDoS attacks, Sybil attacks, Eclipse attacks and possibly others. 

\subsection{Terminology}
In the following we refer to different actors and actions by using the fallowing terminology:
\begin{itemize}
    \item \emph{Crypto-client}: A software illicitly installed in a device belonging to the Corporate Network with the aim of performing the sponge-attack.
    \item \emph{Standard software}: A software legitimately  installed in a device of the Corporate Network.
    \item \emph{Reference device}: A laptop used for running both the Standard software and the Crypto-clients used in this paper.
\end{itemize}

\section{Measurement Setup and Preliminary Considerations}
\label{sec:measurement_setup}

In this section we provide a description of our measurement setup and a preliminary statistical analysis of the collected traces.

\begin{figure}
\includegraphics[width=\columnwidth]{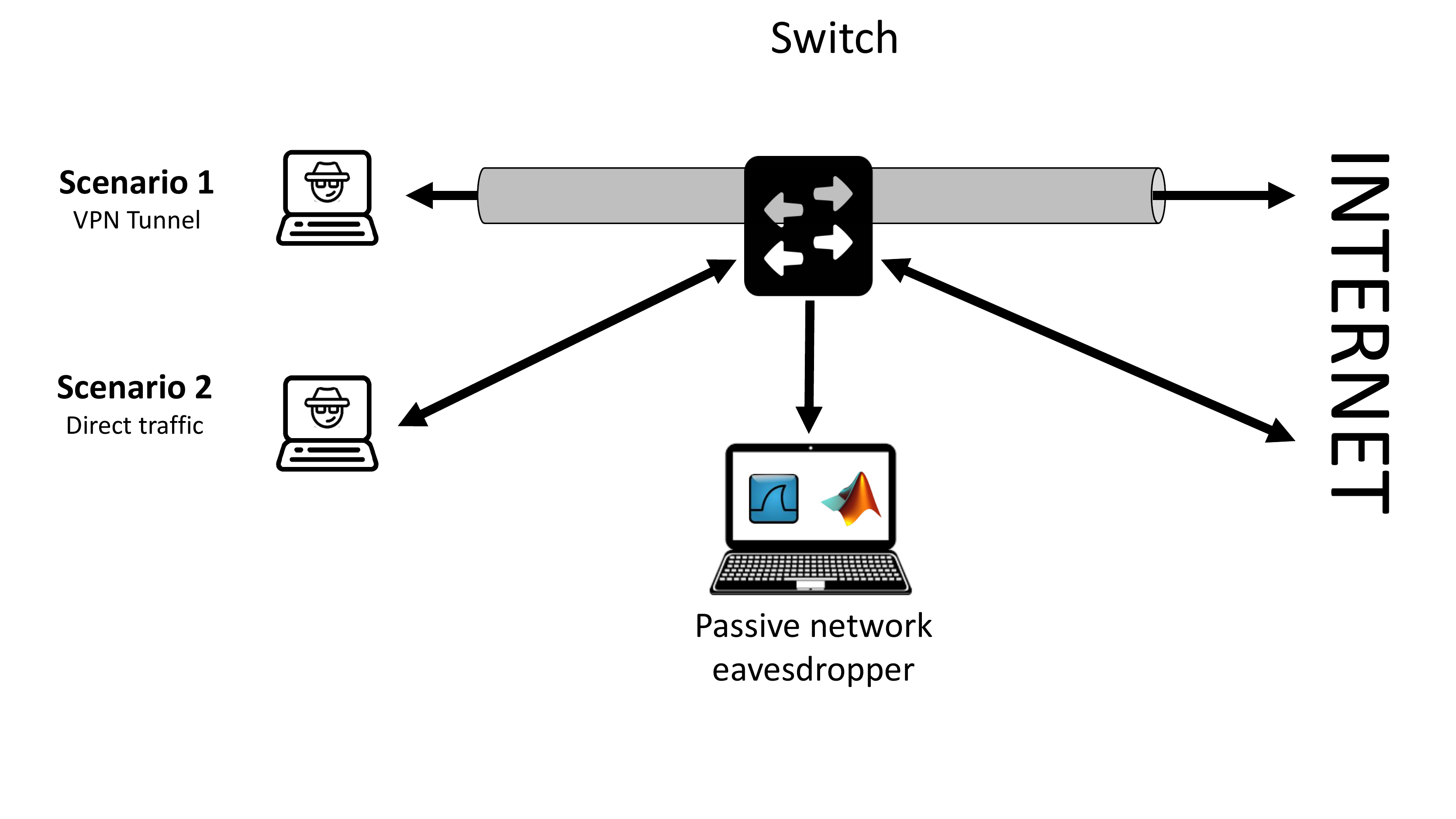}
\centering
\caption{Measurement setup: We consider 2 different scenarios. The malicious device is connected to the network through a VPN Tunnel (Scenario 1) and the malicious device is directly connected to the Internet (Scenario 2). We adopted one laptop for the mining activities (malicious device), one other laptop for collecting all the in-transit packets and finally, a switch featuring a monitoring port.}
\label{fig:measurement_setup}
\end{figure}

{\bf \em Measurement setup.} Our measurement setup can be resumed by Fig.~\ref{fig:measurement_setup}. We consider two scenarios: Scenario 1 where a VPN tunnel adds an encryption layer to the communication, and Scenario 2 where the client is directly connected to the Internet. In Scenario 1, the malicious device is connected to the Internet through an encrypted VPN tunnel. For our measurements, we used two different well-known VPN brands, i.e., Nord VPN (v. 1.2.0) and Express VPN (v. 1.5.0). At the time of writing this paper, Express VPN features more than 2000 servers in 148 countries while Nord VPN features 5064 servers in 62 countries. We arbitrarily set the VPN exit node to France for all our measurements. Conversely, in Scenario 2, the malicious device is directly connected to the Internet without resorting to any additional encryption layer. The malicious device---acting as our reference device when not mining---is a Dell XPS15 laptop running Ubuntu 18.04 (64 bit). All the extracted features are publicly available at~\cite{features}.
\\\\
{\bf Definition.} We define \emph{ingoing flow} all the network traffic from the Internet to our reference device. Moreover, we refer as \emph{outgoing flow} the network traffic generated by the reference device and sent to the Internet.
\\\\
We collected network traffic from three different cryptocurrencies (Bitcoin, Bytecoin and Monero) and three different applications (Skype, YouTube, and standard office applications mixed together) as it follows:

\begin{itemize}
    \item \emph{Skype}. We run an audio Skype-call and collected all the network traffic from/to the reference device.
    \item \emph{YouTube}. We collected the network traffic generated by a random YouTube video from/to the reference device.
    \item \emph{Office network traffic}. We logged the network traffic generated by the reference device while using it for standard office tasks, e.g., e-mail, web-browsing, download and upload of files, Microsoft Office365, etc.
\end{itemize}

The above applications have been selected as a reference excerpt of three traffic patterns coming from three different application scenarios that are audio calls, video streaming, and standard office network traffic. We observe how such network traffic categories cover more than 87\% of the 2018 global consumer internet traffic~\cite{cisco_internet_traffic}. Since our idea is to infer on the presence of a Crypto-client from the network traffic, we considered the network traffic from the above standard applications as the background ``noise" hiding the traffic of the Crypto-client. Our goal is to discriminate the flows involving the Crypto-client from the other flows in the network.

It is known that Machine Learning for network traffic classification is biased by several parameters, i.e., features, type of traffic, trace length, network state, etc. One major concern is related to the consistency of the extracted features given the limited trace length. In particular, we paid particular attention to capture packets from clients at steady-state and after the initial sync period was accomplished. Indeed, Crypto-clients require a warm-up period to download the blockchain and validate it. This guarantees that our log excerpts represent a consistent snapshot of a steady-state client either syncing or mining for the blockchain network.

\begin{table}[]
\centering
\caption{Cryptocurrencies and clients}
\resizebox{0.9\textwidth}{!}{\begin{minipage}{\textwidth}
\bgroup
\def\arraystretch{1.2}
\begin{tabular}{c|c|c|c}
{\bf Cryptocurrency} & {\bf Type} & {\bf Client} & {\bf Version} \\
\hline
Bitcoin & Full Node & Bitcoin Core & 0.17.0 \\
\hline
Bitcoin & Miner & bfgminer & 5.5.0 \\
\hline
Monero  & Full Node & Lithium Luna & 0.12.3.0 \\
\hline
Bytecoin & Full Node & Bytecoin Wallet & 3.3.2 \\
\hline
Bytecoin/Monero & Miner & XMrig & 2.8.1 \\

\hline
\end{tabular}
\egroup
\end{minipage}}
\end{table}

\section{Network Traffic Analysis and Patterns}
\label{sec:network_traffic}

In this section, we start the analysis of the collected network traffic by considering the two network flows: ingoing and outgoing, as explained in the previous section. In order to guarantee a fair comparison between the various scenarios, we extracted the same number of consecutive samples for each network trace, i.e., 4576 samples. Table~\ref{table:traces} shows the network traces we have collected considering the different application scenarios, i.e., Office, Skype, YouTube, Bytecoin, Monero, and Bitcoin. For each scenario, we report the trace duration equivalent to the extracted samples, the quantile 0.5 computed on the interarrival times, and finally the quantile 0.5 computed on the packet sizes. In order to ease the discussion, we refer to each trace by using a sequence of keywords as follows: \emph{[Application][Flow direction][VPN Type]}, where Application can be Office, Skype, YouTube, Bytecoin, Monero, or Bitcoin, Flow direction might be either Ingoing or Outgoing, while VPN Type might be empty (no VPN), Express, or Nord. 

Firstly, we observe how considering the same amount of samples involves very different collection time depending on the application scenario, i.e., about 38.37 seconds for YouTube Ingoing with Express VPN, while about 4598 seconds for Bytecoin Ingoing. In the following we provide some insight from Table~\ref{table:traces}:
\begin{itemize}
    \item {\em Bytecoin.} Interarrival times are significantly affected by the use of VPN, i.e., time reduction spans from 5 to 10 times. Packet sizes are affected as well, i.e., the increase spans from 2 times to 3 times. It is worth noting that the reduction of the interarrival time with the increasing of the packet sizes involves a reduction of the trace length to guarantee the delivery of the same amount of data.
    \item {\em Monero.} VPN tunnelling affects interarrival times of Monero depending on the flow. While ingoing flows experience a reduction of the interarrival time, outgoing flows slightly increase their values. Packet size is affected by the same phenomena. While packet size of ingoing flow ramps up from 66 Bytes (No VPN) to 1433 Bytes (Nord), outgoing flows work in the opposite way decreasing from 1242 Bytes (No VPN) to 131 Bytes (Nord). 
    \item {\em Bitcoin.} Interarrival times are more homogeneous for Bitcoin. Indeed, values span between 180 $\mu s$ and 300 $\mu s$. Nevertheless, we observe that VPN tunnelling affects packet size, indeed for both Nord VPN and Express VPN, packet size is becoming significantly larger.
\end{itemize}

{\bf \em Discussion.} VPN tunnelling tends to squeeze the packets all together and to increase the packet size. Bitcoin is special: VPN tunnelling is affecting much less the original traffic pattern although there are some significant variations for the packet size. It is worth noting the differences among the cryptocurrencies when the traffic is collected without VPN tunnelling. Interarrival times and packet sizes are very different from each other among the currencies as well as between the ingoing/outgoing flows.

\begin{table}[]
\caption{Collected traces: Duration, Median of Interarrival Times, and Median of Packet Sizes.}
\label{table:traces}
\resizebox{0.9\textwidth}{!}{\begin{minipage}{\textwidth}
\begin{tabular}{lllr}
\multicolumn{1}{c}{\textbf{Trace}} & \multicolumn{1}{c}{\textbf{\begin{tabular}[c]{@{}c@{}}Trace\\ duration\end{tabular}}} & \multicolumn{1}{c}{\textbf{\begin{tabular}[c]{@{}c@{}}Int. Time\\ Median\end{tabular}}} & \multicolumn{1}{c}{\textbf{\begin{tabular}[c]{@{}c@{}}Pkt.\\ Size\\ Median\end{tabular}}} \\
 & {[}seconds{]} & {[}seconds{]} & {[}bytes{]} \\
\hline
Office Ingoing & 50.998 & 0.000118 & 1434 \\
Office Outgoing & 59.502 & 0.000311 & 60 \\
Office Ingoing Express   & 84.460 & 0.001003 & 874 \\
Office Outgoing Express   & 147.281 & 0.013116 & 478 \\
Office Ingoing Nord   & 104.700 & 0.000265 & 119 \\
Office Outgoing Nord   & 105.479 & 0.000162 & 1433 \\
\hline
Skype Ingoing & 146.1 & 0.018730 & 136 \\
Skype Outgoing & 145.5 & 0.019988 & 130 \\
Skype Ingoing Express   & 92.577 & 0.020065 & 518 \\
Skype Outoing Express   & 91.911 & 0.019944 & 535 \\
Skype Ingoing Nord   & 87.117 & 0.020119 & 169 \\
Skype Outgoing Nord   & 87.338 & 0.020734 & 196 \\
\hline
YouTube Ingoing & 140.35 & 0.000001 & 1434 \\
YouTube Outgoing & 896.04 & 0.001074 & 54 \\
YouTube Ingoing Express   & 38.378 & 0.001848 & 927.5 \\
YouTube Outgoing Express   & 816 & 0.022749 & 483 \\
YouTube Ingoing Nord   & 168.48 & 0.004814 & 1432 \\
YouTube Outgoing Nord   & 271.93 & 0.007282 & 119 \\
\hline
Bytecoin Ingoing & 4597.2 & 0.004673 & 593 \\
Bytecoin Outgoing & 3280.4 & 0.001130 & 66 \\
Bytecoin Ingoing Express   & 729.67 & 0.000443 & 706 \\
Bytecoin Outgoing Express   & 979.43 & 0.000858 & 134 \\
Bytecoin Ingoing Nord   & 1579 & 0.000803 & 1432 \\
Bytecoin Outgoing Nord   & 2011.1 & 0.000752 & 119 \\
\hline
Monero Ingoing & 822.55 & 0.000450 & 66 \\
Monero Outgoing & 790.58 & 0.000014 & 1242 \\
Monero Ingoing Express   & 197.52 & 0.000044 & 890 \\
Monero Outgoing Express   & 215.45 & 0.000090 & 820 \\
Monero Ingoing Nord   & 445.29 & 0.000137 & 1433 \\
Monero Outgoing Nord   & 404.01 & 0.000117 & 131 \\
\hline
Bitcoin Ingoing & 669.27 & 0.000600 & 90 \\
Bitcoin Outgoing & 659.45 & 0.000242 & 66 \\
Bitcoin Ingoing Express   & 356.1 & 0.000383 & 146 \\
Bitcoin Outgoing Express   & 389.88 & 0.000180 & 146 \\
Bitcoin Ingoing Nord   & 692.44 & 0.000502 & 165 \\
Bitcoin Outgoing Nord   & 822.29 & 0.000359 & 119 \\
\hline
\end{tabular}
\end{minipage}}
\end{table}

\begin{figure}
\includegraphics[width=70mm, scale=0.05]{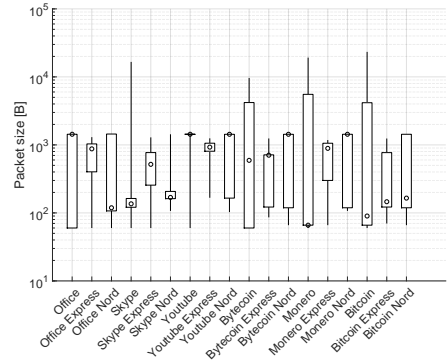}
\centering
\caption{Packet size for outgoing flows: Candle-sticks represent the minimum, quantile 0.05, quantile 0.95 and the maximum packet size for the outgoing flows while the circles represent quantile 0.5.}
\label{fig:packet_size_ingoing}
\end{figure}

\begin{figure}
\includegraphics[width=70mm, scale=0.05]{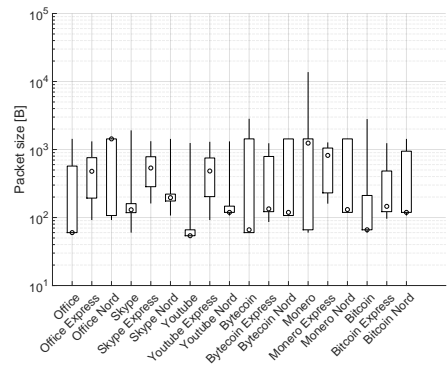}
\centering
\caption{Packet size for ingoing flows: Candle-sticks represent the minimum, quantile 0.05, quantile 0.95 and the maximum packet size for the ingoing flows while the circles represent quantile 0.5.}
\label{fig:packet_size_outgoing}
\end{figure}

Given the above considerations, we consider more in-depth analysis of the flows in order to subsequently identify the features to be used for the Machine Learning process. Figures~\ref{fig:packet_size_ingoing} and~\ref{fig:packet_size_outgoing} show quantile 0.05, 0.5, 0.95, minimum and maximum values associated to each collected trace during our measurements. Outgoing flows present packet sizes very different from each other in the range between 100 and 1000 bytes. Only few exceptions fall out of that range, while it is worth noting how quantile 0.5, e.g., the median, changes for each network trace. Moreover we observe that, raw traffic from Bytecoin, Monero, and Bitcoin present almost the same values of quantile 0.05 and 0.95. Interestingly, such values get closer (being characterized by less variations) when their traffic is tunnelled through a VPN. 

Ingoing flows behave differently from outgoing ones. Packet size spans between closer ranges, i.e., quantile 0.05 and 0.95 are closer with respect to the outgoing flows. 
Median values are randomly distributed and we do not observe any significant pattern in the VPN tunneling of cryptocurrency clients.

\begin{figure}
\includegraphics[width=70mm, scale=0.05]{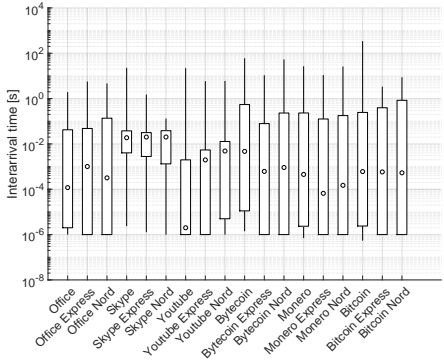}
\centering
\caption{Interarrival times for ingoing flows: Candle-sticks represent the minimum, quantile 0.05, quantile 0.95, and the maximum interarrival times for the ingoing flows---circles representing quantile 0.5.}
\label{fig:int_time_ingoing}
\end{figure}

\begin{figure}
\includegraphics[width=70mm, scale=0.05]{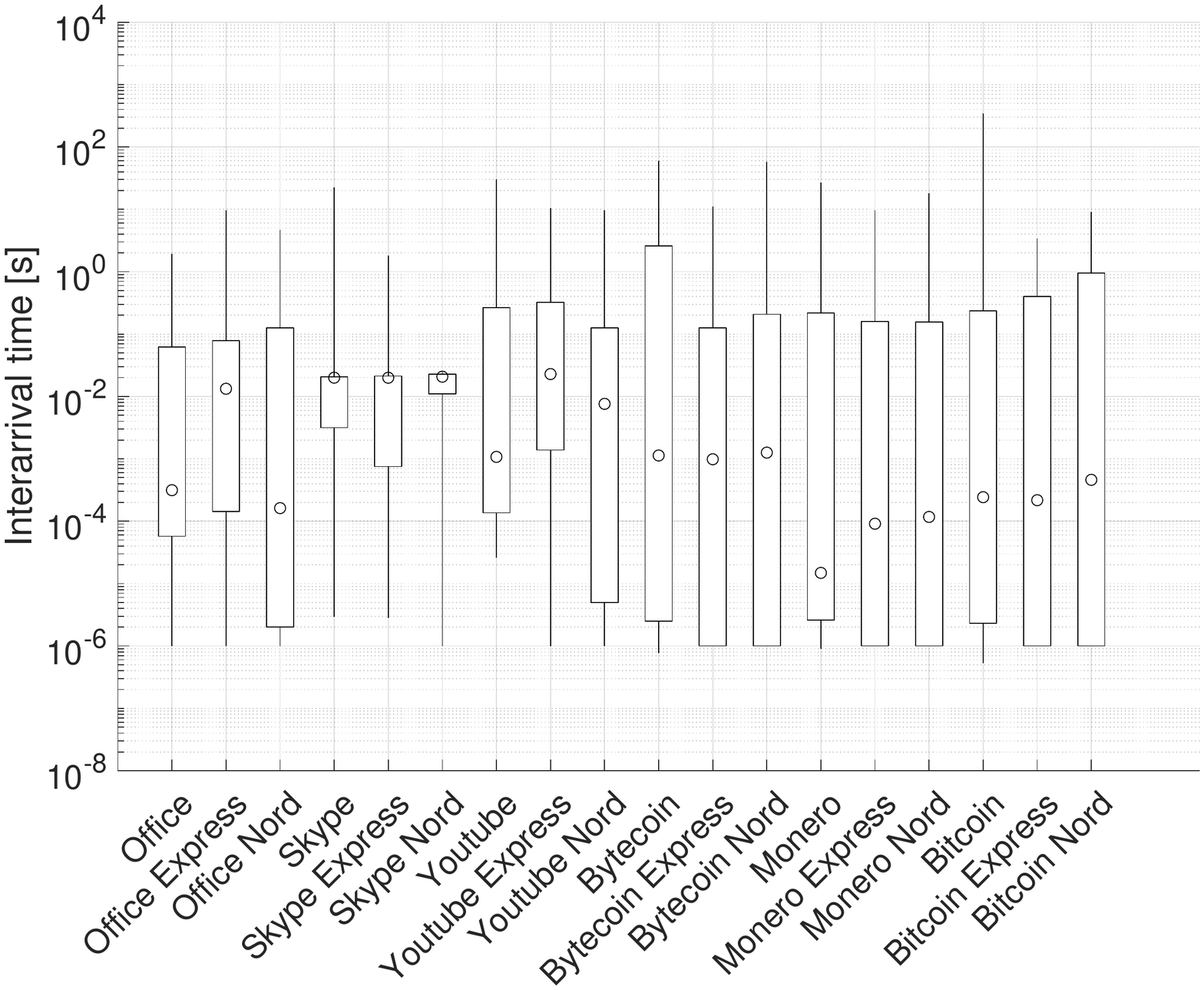}
\centering
\caption{Interarrival times for outgoing flows: Candle-sticks represent the minimum, quantile 0.05, quantile 0.95 and the maximum interarrival times for the outgoing flows while the circles represent quantile 0.5.}
\label{fig:int_time_outgoing}
\end{figure}

We performed the same analysis for the interarrival times obtained by differentiating the absolute arrival times logged by WireShark. The ingoing flows (Fig.~\ref{fig:int_time_ingoing}) of cryptocurrencies are characterized by very similar values, i.e., almost the same quantile 0.05 and 0.95, although we observe that the median values span between $10^{-2}$ and $10^{-4}$ seconds. Similar observations can be drawn by looking at the outgoing flows as depicted by Fig.~\ref{fig:int_time_outgoing}. 

\section{Traffic classification: a baseline example}
\label{sec:traffic_classification}

We implemented all the traffic-classification related tasks in MatLab (R2018a) adopting the Statistics and Machine Learning Toolbox\textsuperscript{\textcopyright}. Our Crypto-client detection algorithm involves the following steps:

\begin{itemize}
    \item {\em Features extraction.} Features identification and extraction are paramount activities to maximize the performance of the classifier. In this work, we consider several features starting from the very standard ones, i.e., interarrival time and packet size. We also consider other derived features with the aim of validating how they affect the final classifier performance. 
    \item {\em k-Fold Cross Validation.} Cross validation is a common practice to average the results of Machine Learning algorithms. It is usually performed by defining a random partition of $k$ out of $n$ observations. The partition divides the observations into $k$ disjoint subsamples (or folds), chosen randomly but with roughly equal size. The default value of $k$ is 10. 
    \item {\em Random Forest (RF).} We adopted the \emph{TreeBagger} MatLab class to implement the RF algorithm. The \emph{TreeBagger} combines the results of many decision trees, which reduces the effects of overfitting and improves generalization. \emph{TreeBagger} grows the decision trees in the ensemble using bootstrap samples of the data. 
    \item {\em Statistics.} This task involves the generation of statistics from the classifier results. Our statistics include (among others) True Negative (TN), False Positive (FP), False Negative (FN), and True Positive (TP), confusion matrix,  etc..
\end{itemize}

In this section, we introduce a simplified version of our methodology considering a binary decision problem. Our goal is to analyze the traffic of a network to determine whether a malicious mining activity is happening. We recall that the traffic collected from the Bitcoin client is related only to the syncing process (being a Full Node), while the one related to the mining process will be considered later on. Moreover, as for the ``noise'' traffic, we adopted a laptop featuring Windows 10 PRO and performing standard office tasks as discussed in the previous section. We now consider only two network traces from Table~\ref{table:traces}: Office Outgoing and Bitcoin Outgoing. Moreover, we assume the hypothesis \emph{$H_0$: the current network event has been generated by the Bitcoin client}. We run the 10-Fold cross validation algorithm using the RF algorithm (with a default value of 20 trees) and only two features: interarrival time and packet size. Table~\ref{table:bitcoin_office} shows the confusion matrix associated to the classifier results, i.e., 4314 times Bitcoin is correctly recognized (True Positive - TP) while 4321 times the class Office is correctly recognized (True Negative - TN). The other values refer to False Positive - FP, i.e., 254 observations are wrongly classified as Bitcoin, and False Negative - FN, i.e., 261 observations are classified as not-Bitcoin (Office) while they actually are. Other interesting metrics---that will be used in the remainder of the paper---are the \emph{True Positive Rate} (TPR) = 0.941, i.e., the number of True Positive normalized to the number of actual Bitcoin observations (TP / (TP + FN)), and the \emph{False Positive Rate} (FPR) = 0.059, i.e., the number of False Positive normalized to the number of predicted observation for Bitcoin (FP / (FP+TN)).

\begin{table}[]
\caption{Baseline example: Bitcoin Vs Office scenario.}
\label{table:bitcoin_office}
\centering
\begin{tabular}{ccc}
& \multicolumn{1}{c|}{\begin{tabular}[c]{@{}c@{}}Predicted\\ No\end{tabular}} & \begin{tabular}[c]{@{}c@{}}Predicted\\ Yes\end{tabular} \\ 
\cline{2-3} 
\multicolumn{1}{c|}{\begin{tabular}[c]{@{}c@{}}Actual\\ No\end{tabular}}  & 4321 & 254 \\ 
\cline{1-1}
\multicolumn{1}{c|}{\begin{tabular}[c]{@{}c@{}}Actual\\ Yes\end{tabular}} & 261  & 4314                                                   
\end{tabular}
\end{table}

\begin{figure}
\includegraphics[width=70mm, scale=0.05]{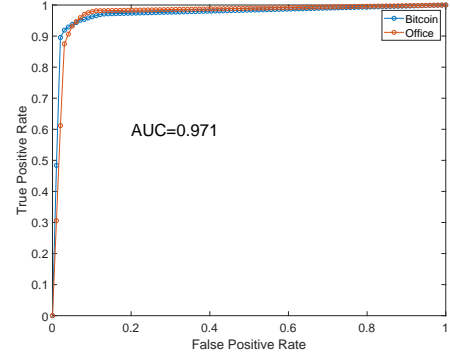}
\centering
\caption{Receiver operating characteristic (ROC) curve: True Positive Rate as a function of the False Positive Rate. The Area Under the Curve (AUC) is about 0.971 for both the application scenarios.}
\label{fig:bitcoin_office_roc}
\end{figure}

FPR and TPR can be used to highlight the classifier performance at different threshold values when the system can accept different levels of false positive values. Figure~\ref{fig:bitcoin_office_roc} shows the Receiver Operating Characteristic (ROC) curve consisting of True Positive Rate (TPR) as a function of False Positive Rate (FPR). Another important metric directly connected to the ROC curve is the so called Area Under the Curve (AUC), i.e., the area under the ROC curve being a value spanning between 0 (worst case) and 1 (best case). As for the ROC curve in Fig.~\ref{fig:bitcoin_office_roc}, AUC is about 0.971 for both the classes, Bitcoin and Office, respectively.

We now add more features to the current scenario and we analyze the performance of the classifier. Let us define the already (basic) introduced features and the new ones, as follows:
\begin{itemize}
    \item Interarrival time ($\delta$): the time elapsed 
    between two 
    consecutive packets. 
    \item Packet Size ($\gamma$): Packet size associated to each packet.
    \item Moving mean of $\delta$ ($\mu_\delta(w)$): each mean value is calculated over a sliding window of length $w$ across neighboring elements of $\delta$.
    \item Moving standard deviation of $\delta$ ($\sigma_\delta(w)$): each standard deviation is calculated over a sliding window of length $w$ across neighboring elements of $\delta$.
    \item Moving mean of $\gamma$ ($\mu_\gamma(w)$): each mean value is calculated over a sliding window of length $w$ across neighboring elements of $\gamma$.
    \item Moving standard deviation of $\gamma$ ($\sigma_\gamma(w)$): each standard deviation is calculated over a sliding window of length $w$ across neighboring elements of $\gamma$.
\end{itemize}

\begin{figure}
\includegraphics[width=\columnwidth]{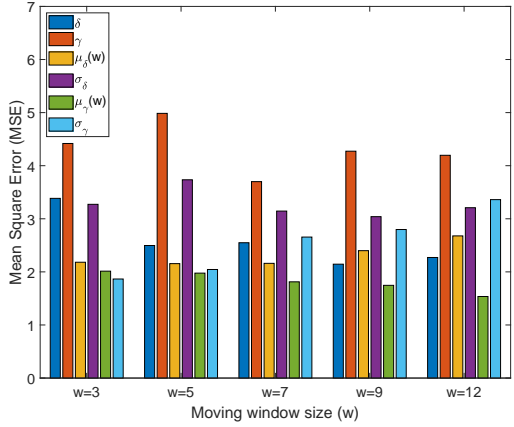}
\centering
\caption{Mean Square Error (MSE) of the classification results as a function of the features and the moving window size ($w$).}
\label{fig:bitcoin_office_w}
\end{figure}

In order to evaluate the impact of the features on the classification algorithm we used the Mean Square Error (MSE) averaged over all the trees in the ensemble and divided by the standard deviation taken over the trees, for each feature. The larger this value, the more important the feature is in the classification process. Figure~\ref{fig:bitcoin_office_w} shows the Mean Square Error (MSE) as a function of the moving window size ($w$) and the different type of features. Firstly we observe that, for this scenario---Bitcoin Vs Office---the most important feature is $\gamma$, i.e., the packet size, represented by the red bar. The other features have about the same weights, while it turns out that $w=5$ is  a good trade-off for the window size of the moving mean and the standard deviation.

\section{Crypto-Aegis: Detection and Identification of Full Nodes}
\label{sec:detection_full_nodes}
In this section, we consider the traces of Table~\ref{table:traces} while  parting them into ingoing and outgoing flows. As previously discussed, we consider three main metrics: True Positive Rate (TPR), False Positive Rate (FPR) and the Area Under the Curve (AUC). Our RF classifier has been configured with 20 default trees, the 6 features already introduced in the previous section, and a moving window of 5 observations. Moreover, we consider only Full Node clients; therefore, the observed network traffic will be related to syncing and consensus operations. 

\begin{figure}
\includegraphics[width=\columnwidth]{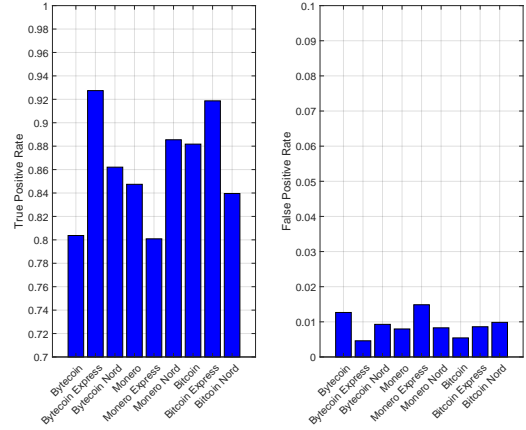}
\centering
\caption{True Positive Rate and False Positive Rate for ingoing network traffic of a Full Node considering different cryptocurrencies.}
\label{fig:ingoing_fpr_tpr}
\end{figure}

{\bf \em Ingoing flows.} Figure~\ref{fig:ingoing_fpr_tpr} shows TPR and FPR for all the cryptocurrencies we have considered in this work. Firstly, we observe how the overall results are quite satisfactory, i.e., the mean computed on the TPR and FPR values is about 0.86 and 0.0088, respectively. The best detection performance are achieved over Bytecoin Express (TPR=0.92, FPR=0.0047) and Bitcoin Express (TPR=0.92, FPR=0.008). Conversely, worst case performance are achieved for:
\begin{itemize}
    \item Bytecoin (TPR=0.81, FPR=0.012). Misclassifications are mainly due to Monero (332 cases - 7\%),  Bitcoin (106 cases - 2\%), and Bytecoin Nord (97 cases - 2\%).
    \item Monero Express (TPR=0.80, FPR=0.014). False positive are mainly due to Office Express (381 case - 8\%), YouTube Express (148 cases - 3\%) and Bytecoin Express (63 cases - 1\%).
    \item Bitcoin Nord (TPR=0.84, FPR=0.009). Classification errors mainly come from Monero Nord (251 cases - 5\%), Bitcoin Express (217 cases - 4\%) and Bytecoin Nord (70 cases - 1\%).
\end{itemize}

Our results prove that ingoing flows (from the Internet to the device) can be used to effectively identify malicious miners inside local networks. In particular, traffic generated by cryptocurrencies clients (without VPN) can be detected with high TPR values, i.e., 0.84, 0.87 and 0.90 for Bytecoin, Monero, and Bitcoin. The adoption of a VPN tunnel does not improve the privacy of the Crypto-client: TPR is increasing for Bytecoin when tunneled through a VPN, while Monero and Bitcoin have diverging performance as a function of the adopted VPN brands. Moreover, we observe that worst case performance are due to Crypto-clients misclassified for other Crypto-clients; indeed, there is only one exception: 148 cases of YouTube Express classified as Monero Express. We consider the previous phenomenon as a by-product of the VPN tunnelling; indeed, it is reasonable to assume that the same VPN is (slightly) re-shaping different traffic patterns in the same way.

\begin{figure}
\includegraphics[width=\columnwidth]{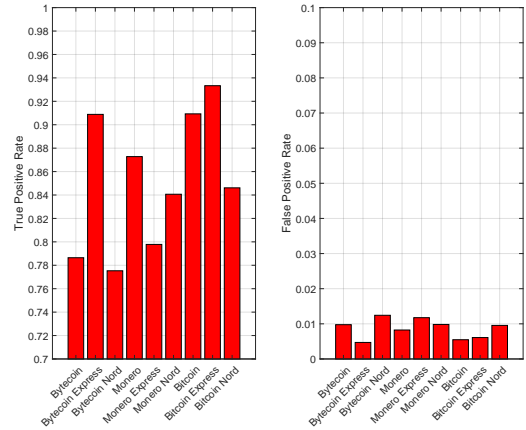}
\centering
\caption{True Positive Rate and False Positive Rate for outgoing network traffic of a Full Node considering different cryptocurrencies.}
\label{fig:outgoing_fpr_tpr}
\end{figure}

{\bf \em Outgoing flows.} Figure~\ref{fig:outgoing_fpr_tpr} shows TPR and FPR for all the cryptocurrencies we have considered in this work. As for the ingoing flows, we observe how the overall results are, again, quite satisfactory, i.e., the mean value computed on the TPR and the FPR values is 0.85 and 0.008, respectively. The best detection performance are achieved over Bitcoin Express (TPR=0.93, FPR=0.006) and Bitcoin (TPR=0.90, FPR=0.005). Conversely, worst case performance are achieved for:
\begin{itemize}
    \item Monero Express (TPR=0.79, FPR=0.011). Misclassifications are mainly due to Office Express (563 cases - 12\%) and YouTube Express (295 cases - 6\%).
    \item Bytecoin Nord (TPR=0.77, FPR=0.012). False positive are mainly due to Monero Nord (281 cases - 6\%) and Bitcoin Nord (202 cases - 4\%).
    \item Bytecoin (TPR=0.78, FPR=0.009). Classification errors mainly come from Monero (345 cases - 7\%) and Bitcoin (147 cases - 3\%).
\end{itemize}

The above considerations prove that outgoing flows (from the device to the Internet) can be used to effectively identify malicious miners inside local networks. Crypto-clients (without VPN) can be detected with TPR of about 0.80 (0.78), 0.84 (0.87) and 0.88 (0.90), for Bytecoin Ingoing (Outgoing), Monero Ingoing (Outgoing) and Bitcoin Ingoing (Outgoing), respectively. As for the ingoing flow analysis, VPN does not significantly increase the privacy of the device: TPR values fluctuate depending on the adopted VPN brand, but they still remain high---and even increase for some cases. 
One more interesting aspect is that misclassifications happen among cryptocurrencies adopting the same VPN, i.e., a Crypto-client is misclassified as belonging to another cryptocurrency but still using the same VPN; although this is a marginal effect (most frequent case is less than 12\%) we can reasonably suspect that  VPN tunnelling is shaping the traffic patterns; that is why (at least partially) the Machine Learning algorithm experience such type of misprediction.

\begin{figure}
\includegraphics[width=70mm, scale=0.05]{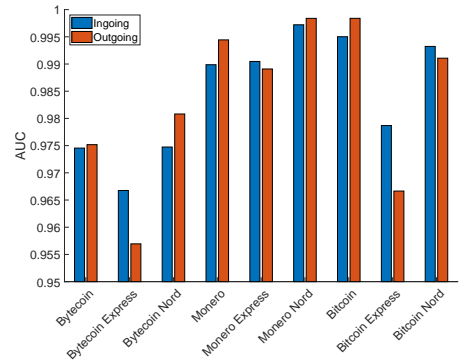}
\centering
\caption{Area Under the Curve (AUC) for both ingoing and outgoing flows of a Full Node considering different cryptocurrencies.}
\label{fig:auc_inout}
\end{figure}

{\bf \em Area Under the Curve (AUC).} As previously introduced, the Area Under the Curve (AUC) is the area under the ROC curve. Its value spans between 0 and 1, and the largest it is, the better the classifier's performance are. Figure~\ref{fig:auc_inout} shows the AUC values for the different traces/scenarios in Table~\ref{table:traces}. For both the ingoing and outgoing flows, the worst performance are experienced by Bytecoin Express, Bitcoin Express, and Bytecoin. Of course, this does not imply that VPN tunnelling is protecting the privacy of the clients as previously discussed. Indeed, we do not observe any significant difference with or without the presence of a VPN tunnel: for all the three cryptocurrencies the AUC values in presence of a VPN tunnel are still very high, i.e., larger than 0.955.

\section{Crypto-Aegis: Detection and Identification of Miners}
\label{sec:detection_miners}

In the following, we consider cryptocurrency clients acting as pure Miners as already introduced in Section~\ref{sec:scenario}. Therefore, we consider \emph{bfgminer} for both Bitcoin and \emph{XMrig} for Bytecoin and Monero as illicitly installed clients in a device belonging to a Corporate Network. As for the full node's case, we consider two different VPN, i.e., Express VPN and Nord VPN, and two flows, i.e., ingoing and outgoing.

In the vast majority of cases, the mining task is performed by mining pools~\cite{bonneau}~\cite{rosenfeld}. In practice, Miners collaborate in pools to lower the variance of their revenue by sharing rewards with a group of other Miners. We consider the same measurement setup of Fig.~\ref{fig:measurement_setup}, but we used a Miner as the peer for the blockchain network. As for the analysis of the Full Nodes (Section~\ref{sec:detection_full_nodes}), we part all the collected traces into two subsets: ingoing and outgoing flows.

Table~\ref{table:mining_traces} resumes the network traces we have collected when considering a Miner as a client: for each trace, we report the duration, the median of both the interarrival times, and the packet sizes. Firstly, we reduce all the traces to the same amount of packets to guarantee a fair classification process for the RF algorithm: we choose 832 packets. We observe how the number of packets from the Miners is less than the one of the Full Nodes: this is mainly due to the fact that Miners generate much less traffic, as it turns out by comparing the interarrival times from Table~\ref{table:mining_traces} with the ones from Table~\ref{table:traces}. Moreover, it is worth noting the case of Monero: the interarrival times are decreasing from 13.97 (Ingoing) and 6.11 (outgoing) seconds to less than 3 seconds (Express), or 1 second (Nord) when VPN tunneling is activated. Conversely, VPN does not significantly affect packet size with one exception: Monero Ingoing Nord experiences a packet size of 428 Bytes, while the packet size of Monero Outgoing with no VPN is 66 Bytes.

\begin{table}[]
\centering
\caption{Collected traces from Miners: Duration, Median of Interarrival Times, and Median of Packet Sizes.}
\label{table:mining_traces}
\begin{adjustbox}{max width=\columnwidth}
\begin{tabular}{llrr}
\multicolumn{1}{c}{\textbf{Trace}} & \multicolumn{1}{c}{\textbf{\begin{tabular}[c]{@{}c@{}}Trace\\ duration\end{tabular}}} & \multicolumn{1}{c}{\textbf{\begin{tabular}[c]{@{}c@{}}Int. Time\\ Median\end{tabular}}} & \multicolumn{1}{c}{\textbf{\begin{tabular}[c]{@{}c@{}}Pkt.\\ Size\\ Median\end{tabular}}} \\
 & {[}seconds{]} & {[}seconds{]} & {[}bytes{]} \\
\hline
Bytecoin Ingoing              & 3935.21  & 2.41 & 131 \\
Bytecoin Outgoing             & 2558.00  & 0.39 & 66  \\
Bytecoin Ingoing Express  & 2486.86  & 1.47 & 187 \\
Bytecoin Outgoing Express & 1961.86  & 0.27 & 122 \\
Bytecoin Ingoing Nord     & 2513.18  & 0.37 & 184 \\
Bytecoin Outgoing Nord    & 2142.08  & 0.29 & 119 \\
\hline
Monero Ingoing                & 15377.52 & 13.97 & 131 \\
Monero Outgoing               & 10483.16 & 6.11  & 66  \\
Monero Ingoing Express    & 3658.87  & 2.46  & 173 \\
Monero Outgoing Express   & 3688.58  & 2.53  & 122 \\
Monero Ingoing Nord       & 6184.70  & 0.34  & 429 \\
Monero Outgoing Nord      & 5218.86  & 0.17  & 119 \\
\hline
Bitcoin Ingoing               & 8075.92  & 0.29 & 125 \\
Bitcoin Outgoing              & 8014.65  & 0.37 & 66  \\
Bitcoin Ingoing Express   & 2315.13  & 0.12 & 181 \\
Bitcoin Outgoing Express  & 2797.90  & 0.37 & 122 \\
Bitcoin Ingoing Nord      & 4756.34  & 0.13 & 178 \\
Bitcoin Outgoing Nord     & 4644.65  & 0.19 & 119 \\
\hline
\end{tabular}
\end{adjustbox}
\end{table}

As for the previous case, we consider True Positive Rate (TPR) and False Positive Rate (FPR) as our reference metrics. 

\begin{figure}
\includegraphics[width=\columnwidth]{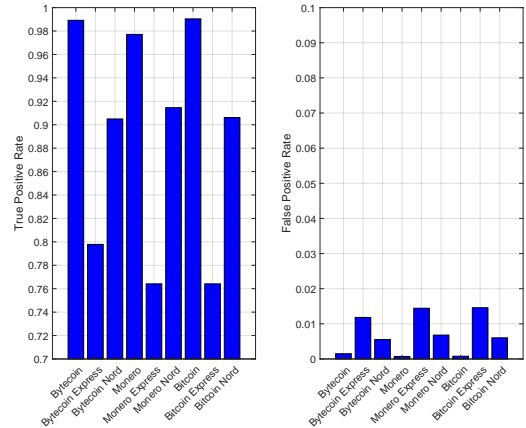}
\centering
\caption{True Positive Rate and False Positive Rate for ingoing network traffic of Miners of different cryptocurrencies.}
\label{fig:mining_ingoing_fpr_tpr}
\end{figure}

{\bf Ingoing flows.} Figure~\ref{fig:mining_ingoing_fpr_tpr} shows the TPR and the FPR for different Miner clients when their ingoing flow is compared with the ingoing flows of Office, Skype, and YouTube (same traces from the Standard software of Table~\ref{table:traces}). Firstly we observe how raw traffic (not  tunneled through a VPN) is better identified: Bytecoin, Monero, and Bitcoin turn out to have TPR values equal to 0.993, 0.983, and 0.985, respectively, while FPR values are equal to 0.0018, 0.0004, and 0.0011, respectively. 

Worst case performance are due to Bytecoin Express, Monero Express and Bitcoin Express:
\begin{itemize}
    \item Bytecoin Express (TPR = 0.79, FPR = 0.011). Misclassifications are mainly due to Monero Express (36 cases - 4\%) and Bitcoin Express (10 cases - 1\%).
    \item Monero Express (TPR = 0.77, FPR = 0.015). False positive are mainly due to Bytecoin Express (56 cases - 6\%).
    \item Bitcoin Express (TPR = 0.77, FPR = 0.013). Classification errors mainly come from Monero Express (102 cases - 12\%) and Bytecoin Express (68 cases - 8\%).
\end{itemize}

\begin{figure}
\includegraphics[width=\columnwidth]{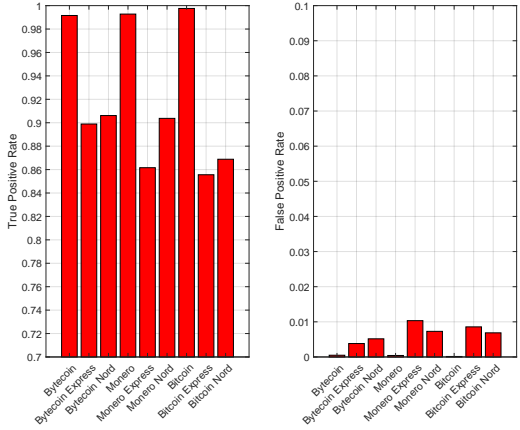}
\centering
\caption{True Positive Rate and False Positive Rate for outgoing network traffic of Miners of different cryptocurrencies.}
\label{fig:mining_outgoing_fpr_tpr}
\end{figure}

{\bf Outgoing flows.} Figure~\ref{fig:mining_outgoing_fpr_tpr} shows TPR and FPR associated to the outgoing flows for all the considered Miner clients compared with Office, Skype, and YouTube network traces. The mean values associated to TPR and FPR are 0.946 and 0.003. 
The best detection performance are achieved with Bytecoin, Monero, and Bitcoin without VPN tunneling as in the previous case (Ingoing flow analysis). As in this case (ingoing flows), VPN tunneling affects the privacy of the client decreasing the identification performance, and in the following, we comment on the worst cases:

\begin{itemize}
    \item Monero Express (TPR=0.86, FPR=0.009). Misclassifications are mainly due to Bitcoin Express (53 cases - 6\%) and Bytecoin Express (20 cases - 2\%).
    \item Bitcoin Express (TPR=0.86, FPR=0.008). False positive are mainly due to Monero Express (53 cases - 6\%) and Bytecoin Express (13 cases - 1\%).
\end{itemize}

We highlight how---for both the Outgoing and Ingoing cases--- misclassification are only due to cryptocurrencies, i.e., the vast majority of the False Positive are due to other cryptocurrencies and not to Standard software traffic. Finally, it is interesting to observe how VPNs are more effective to protect the privacy of the Miners respect to the Full Node scenario. This is mainly due to the fact that the overall traffic sent/received by each Miner is significantly less than that one of Full Nodes and it is buried in the VPN-encrypted traffic. 

\section{Crypto-Aegis: Sponge-attack detection}
\label{sec:sponge_attack_detection}

In this section, we consider the more general binary decision problem of detecting the presence of a Crypto-client in the scenario of Fig.~\ref{fig:scenario}. Therefore, we assume the hypothesis \emph{$H_0$: the current network event has been generated by a Crypto-client}. We consider the traces from the Standard software previously introduced, i.e., Office, Skype, and YouTube, and we test them against all the traces involving a Crypto-clients. As a preliminary step, we pre-process all the traces and uniform their lengths to obtain two classes: the first one constituted by Office, Skype, and YouTube traces (evenly distributed); and, the second one constituted by all the traces from the Crypto-clients (Bytecoin, Monero, and Bitcoin) with and without the VPN tunnels. 
Table~\ref{table:ingoing_sponge_attack} (~\ref{table:outgoing_sponge_attack}) shows the confusion matrix associated to the previously introduced binary classifier assuming the ingoing (outgoing) network traffic. Firstly, we observe that True Positive sums up to 290728, i.e., the number of times Crypto-clients are correctly identified. Moreover, the classifier correctly identifies 288452 events as network traffic from Standard software. Nevertheless, there are 10255 and 7979 false positive and false negative events, respectively. The results on the outgoing traffic are characterized by similar outstanding performance, i.e., TP=276187, TN=275969, FP=9370, and FN=9152.

\begin{table}[]
\caption{Sponge-attack detection: Ingoing Traffic Classification.}
\label{table:ingoing_sponge_attack}
\centering
\begin{tabular}{ccc}
& \multicolumn{1}{c|}{\begin{tabular}[c]{@{}c@{}}Predicted\\ No\end{tabular}} & \begin{tabular}[c]{@{}c@{}}Predicted\\ Yes\end{tabular} \\ 
\cline{2-3} 
\multicolumn{1}{c|}{\begin{tabular}[c]{@{}c@{}}Actual\\ No\end{tabular}}  & 288452 & 10255 \\ 
\cline{1-1}
\multicolumn{1}{c|}{\begin{tabular}[c]{@{}c@{}}Actual\\ Yes\end{tabular}} & 7979  & 290728                                                   
\end{tabular}
\end{table}

\begin{table}[]
\caption{Sponge-attack detection: Outgoing Traffic Classification.}
\label{table:outgoing_sponge_attack}
\centering
\begin{tabular}{ccc}
& \multicolumn{1}{c|}{\begin{tabular}[c]{@{}c@{}}Predicted\\ No\end{tabular}} & \begin{tabular}[c]{@{}c@{}}Predicted\\ Yes\end{tabular} \\ 
\cline{2-3} 
\multicolumn{1}{c|}{\begin{tabular}[c]{@{}c@{}}Actual\\ No\end{tabular}}  & 275969 & 9370 \\ 
\cline{1-1}
\multicolumn{1}{c|}{\begin{tabular}[c]{@{}c@{}}Actual\\ Yes\end{tabular}} & 9152  & 276187                                                   
\end{tabular}
\end{table}

\begin{figure}
\includegraphics[width=70mm, scale=0.05]{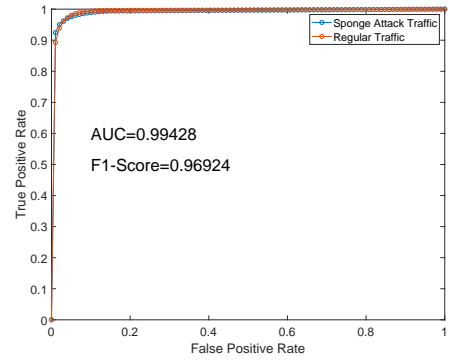}
\centering
\caption{Sponge-attack detection for Ingoing flows: True Positive Rate as a function of the False Positive Rate assuming the traffic as generated by two classes: Crypto-clients (Miners and Full Nodes) Vs standard applications.}
\label{fig:sponge_attack_roc_ingoing}
\end{figure}

Figure~\ref{fig:sponge_attack_roc_ingoing} shows the Receiver Operating Characteristic (ROC) curve consisting of the True Positive Rate (TPR) as a function of the False Positive Rate (FPR) assuming only the ingoing flows. We also report: the Area Under the Curve (AUC), being it equal to about 0.99428 for the ingoing traffic; and, the F1-score --- given by the harmonic average of the \emph{precision} (TP / (TP + FP)) and the \emph{recall} (TP / (TP + FN)). F1-score is a measure of the accuracy: it is  equal to 1 when there is perfect both precision and recall, while it approaches  0 when  conditions worsen.

\begin{figure}
\includegraphics[width=70mm, scale=0.05]{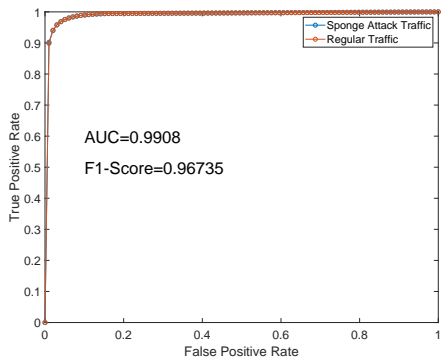}
\centering
\caption{Sponge-attack detection for Outgoing flows: True Positive Rate as a function of the False Positive Rate assuming the traffic being generated by two classes: Crypto-clients (Miners and Full Nodes) Vs standard applications.}
\label{fig:sponge_attack_roc_outgoing}
\end{figure}

Figure~\ref{fig:sponge_attack_roc_outgoing}  shows the Receiver Operating Characteristic (ROC) curve consisting of the True Positive Rate (TPR) as a function of the False Positive Rate (FPR), assuming only the outgoing flows. As for the previous case, we report both the AUC and the F1-score. 

{\bf Time to detect the Crypto-client.} Recalling Fig.~\ref{fig:bitcoin_office_w}, we observe that the detection time strictly depends on the number of packets requested by the feature generation algorithm; in particular, the moving mean and the moving standard deviation. Full Nodes are characterized by very short interarrival times, i.e., worst case is equal to 0.022749 seconds (median value for YouTube Outgoing Express), and therefore, waiting for 5 subsequent packets (moving window size used throughout this paper) involves a really short period of time, i.e., less than 120 ms. 
Even the longest window size depicted in Fig.~\ref{fig:bitcoin_office_w}, i.e., 12 packets, requires less than 300ms. Conversely, Miners require more time for being identified due to their stretched interarrival times. We observe that there are specific Crypto-client configurations from Table~\ref{table:mining_traces} that are characterized by large interarrival times, i.e., Bytecoin Ingoing and Monero Ingoing. For such cases, assuming a moving window size $w$ equal to 5 consecutive packets, 12.05 and 69.85 seconds are required, respectively.

\section{Discussion and Comparison with other solutions}
\label{sec:discussion}

In this section, we discuss the results shown in  this paper, while also  comparing our Crypto-Aegis framework  against competing solutions existing in the literature.

{\bf Full Node.} Our solution turned out to be very effective for the detection and identification of Full Nodes in local networks. The RF algorithm is able to independently detect and identify Full Nodes by leveraging either ingoing and outgoing flows. TPR and FPR metrics have similar values although we observe a few exceptions related to the worst cases: Bytecoin Ingoing and Monero Express Ingoing (TPR $>$ 0.8), and Bytecoin Outgoing and Bytecoin Nord Outgoing (TPR $>$ 0.79). Conversely, FPR are always less than 0.015, guaranteeing an extremely low number of potential false alarms. 

{\bf Miner.}  Miners' detection performance are similar to the previous case, although we observe that TPR values for the outgoing flows are (in general) greater than the ones for the ingoing flows. This is mainly due to worse anonymization performance achieved by Express VPN for all the ingoing flows. More in general, we highlight the outstanding performance of the classifier when the crypto-clients' network traffic is not tunneled through a VPN. Finally, we stress that, for both the above scenarios (Full Nodes and Miners), false positives are mainly due to cryptocurrencies predicted as other cryptocurrencies. This significantly mitigates the more general problem of the number of false negative when considering the detection of a Crypto-client at large: since FP are still due to other cryptocurrencies, the detection algorithm could be still considered successful (although not being able to fully identify the Crypto-client).

To the best of our knowledge, our contribution is the first one to leverage Machine Learning techniques to detect Crypto-clients by analyzing the network traffic---though ML techniques have been extensively used to effectively detect anomalies in network traffic such as malware, intrusion detection, etc.. In this paper we have proved that network traffic generated by Crypto-clients is characterized by pattern anomalies with respect to standard applications such as Skype, YouTube, and standard software used during typical office tasks. 

Naive solutions~\cite{CoinBlockerLists}~\cite{drmine}~\cite{MinerBlock} dealing with pure web-based mining (cryptojacking) involve blacklisting, a set of malicious URLs reported as suspicious from different sources, i.e., Twitter, blogs, etc. The idea mainly resorts to the installation of a plugin for the web-browser and by monitoring the connections, preventing those belonging to the blacklist. Blacklisting is neither effective nor efficient since it suffers from many false negatives due to URL randomization and requires the installation of third-party software in all the corporate devices. Other solutions involve to monitor the CPU throttle and asking for extra permissions to the web-browser when a process requires high CPU usage~\cite{troy}. While monitoring the CPU is a promising solution, such a technique potentially suffers from high false positive rates due to the difficulty of discriminating between the miners and other CPU demanding processes, such as videogames. 
Another interesting technique has been recently proposed by~\cite{konoth2018minesweeper}: authors combined multiple techniques to eventually identifying cryptographic operations and inferring on the execution of a miner.

We observe that while the above solutions can be adopted for the detection of Miners under certain conditions, they do not take into account Full Node detection. For this reason, the \textit{solo} mining activity, as well as other types of attacks described in Section~\ref{sec:adversarial_model}, are not recognized by existing solutions. Moreover, none of these solutions can be used to identify the cryptocurrency used by the attacker.
Finally, we highlight that all the host-based solutions can be only adopted for the detection of outsider adversaries delivering the crypto-jacking attacks to the users browsing the web. Indeed, none of the above solutions can be used for detecting insider adversaries, i.e., malicious corporate employee willing to run the attack from inside the Corporate Network by having full rights and control of their devices, i.e., laptops, desktops, servers, etc..
Table~\ref{table:comparison} wraps up on the comparison between our solution and the existing ones from the literature.

\begin{table}
\footnotesize
\caption{Comparison between our solution and the ones  in the literature. \label{table:comparison}}
\resizebox{0.9\textwidth}{!}{\begin{minipage}{\textwidth}
\begin{tabular}{c|c|c|c|c|c|c}
  & \multicolumn{2}{c|}{Detection} & \multicolumn{2}{c|}{Identification} & \multicolumn{2}{c}{\begin{tabular}[c]{@{}c@{}}Adversary\\ Model\end{tabular}} \\ 
  & \multicolumn{1}{c|}{\begin{tabular}[c]{@{}l@{}}Full \\ Node\end{tabular}} & \multicolumn{1}{c|}{Miner} & \multicolumn{1}{c|}{\begin{tabular}[c]{@{}l@{}}Full \\ Node\end{tabular}} & \multicolumn{1}{c|}{Miner} & \multicolumn{1}{c|}{\rotatebox{45}{Insider}} & \multicolumn{1}{c}{\rotatebox{45}{Outsider}} \\ \hline
\makecell[c]{Blacklisting\\~\cite{CoinBlockerLists}~\cite{drmine}~\cite{MinerBlock}} & \xmark & \cmark & \xmark & \xmark & \xmark & \cmark \\ \hline
\begin{tabular}[c]{@{}c@{}} CPU-throttle\\ monitoring\\~\cite{troy}\end{tabular} & \xmark & \cmark & \xmark & \xmark & \xmark & \cmark \\ \hline 
\makecell[c]{MineSweeper\\~\cite{konoth2018minesweeper}} & \xmark & \cmark & \xmark & \xmark & \xmark & \cmark \\ \hline
\makecell[c]{\colorTextMod{CapJack}\\~\cite{8737381}} & \xmark & \cmark & \xmark & \xmark & \xmark & \cmark \\ \hline
\makecell[c]{\colorTextMod{CMTracker}\\~\cite{hong2018you}} & \xmark & \cmark & \xmark & \xmark & \xmark & \cmark \\ \hline
\makecell[c]{Crypto-Aegis} & \cmark & \cmark & \cmark & \cmark & \cmark & \cmark  
\end{tabular}
\end{minipage}}
\end{table}

\section{Conclusion}
\label{sec:conclusion}
In this paper, in the context of  unauthorized crypto-mining activities, we have first proposed a novel attacker model ({\em sponge-attack}) that subsumes the attacker model present in the literature (cryptojacking). Then, we have introduced Crypto-Aegis, a ML based framework that is able to detect and identify crypto-mining activities related to the sponge-attack. Crypto-Aegis enjoys several features;  
in particular:  (i) it is infrastructure  independent;  (ii) it is device  independent; (iii) it supports multi-adversarial profiles; (iv) it does not require a  clean  state  to operate; and,  (v) it is highly effective (e.g., F1-score of 0.96 and an AUC for the ROC greater than 0.99). Moreover, we have also proved that the Crypto-Aegis framework is resilient to the adoption (by the adversary) of VPN tunnelling. The quality and viability of the achieved results, superior to competing solutions in the literature, combined with 
the novelty of the introduced attacker model, pave the way for further research in this domain.

\section*{Acknowledgements}
This publication was partially supported by awards NPRP11S-0109-180242, UREP23-065-1-014, and NPRP X-063-1- 014 from the QNRF-Qatar National Research Fund, a member of The Qatar Foundation. The information and views set out in this publication are those of the authors and do not necessarily reflect the official opinion of the QNRF.




\balance

\bibliographystyle{cas-model2-names}

\bibliography{cryptojacking}



\end{document}